\documentstyle[12pt,epsf]{article}

\def\sla  {\!\!\!\slash}

\def\ds {\displaystyle}

\newcommand{\be}{\begin{equation}}
\newcommand{\ee}{\end{equation}}
\newcommand{\bea}{\begin{eqnarray}}
\newcommand{\ena}{\end{eqnarray}} 

\newcommand{\np}{Nucl.\,Phys.\,}
\newcommand{\pl}{Phys.\,Lett.\,}
\newcommand{\pr}{Phys.\,Rev.\,}
\newcommand{\prl}{Phys.\,Rev.\,Lett.\,}

\newcommand{\eps}{\varepsilon}

\newcommand{\ra}{\rightarrow}

\newcommand{\ie}{{\em i.e.\ }}

\begin{document}
\bibliographystyle{unsrt}
\begin{titlepage}
\pagestyle{empty}
\begin{center}
{\bf\Huge Laboratoire d'Annecy-le-Vieux de }\\
\vskip 2mm
{\bf\Huge Physique Th\'eorique}
\end{center}
\begin{center}
\rule{16cm}{.5mm}
\end{center}
\relax

\def\baselinestretch{1.2}
\vspace*{\fill}
\begin{center}
{\large {\bf Improvement of the Calculation of Scattering Amplitudes 
with External Fermions}}

\vspace*{0.5cm}

 E.~Chopin$^{1}$\\ 

{\it Laboratoire d'Annecy-le-Vieux de Physique Th\'eorique 
LAPTH.}\footnote{ URA 14-36 du CNRS, associ\'ee \`a l'Universit\'e de 
Savoie.}\\ 
{\it B.P.110, 74941 Annecy-Le-Vieux Cedex, France} \\
{\tenrm 1. e-mail:chopin@lapp.in2p3.fr} \\
{\tenrm PACS numbers: 11.80-Cr, 11.30-Cp, 11.55-m}
\end{center}
\vspace*{\fill}

\centerline{ {\bf Abstract} }
\baselineskip=14pt
\noindent
 {\small 
In this paper, we present an improvement of a method for 
computing scattering amplitudes that include external 
(polarized) fermions with the following features: the formulas are  
quite general and work for different kinematic configurations and 
different external masses, they are explicitly covariant, they do 
not depend on a specific representation of the Dirac matrices and 
they have a meaningful limit when the masses tend to $0$. 
The results presented make use of some well known formulas 
describing the density matrices in terms of projection operators 
within a more general formalism. Since our formulas intend to 
be as general as possible, we also take into account the 
possibility of a transverse polarization for massless fermions. 
 }
\vspace*{\fill}
 
\vspace*{0.1cm}
\rightline{LAPTH-682/98}

\end{titlepage}
\baselineskip=16pt
\def\baselinestretch{1.1}
\input{FEYNMAN}
\bigphotons

  The notion of spin is somewhat difficult to express in a general, 
covariant formalism which would have also a meaningful limit for 
$m\ra 0$, since the Poincar\'e little groups are different for 
$p^2>0$ and $p^2=0$. Therefore, for the computation of Feynman 
amplitudes, we cannot obtain general formulas that are exactly 
continuous in the limit $m\ra 0$. Furthermore, when fermions are 
involved, one has to deal with spinors and Dirac matrices, 
and no particular representation of these matrices has some physical 
meaning. But for practical calculations involving spinors, a lot of 
people still make reference to a specific representation 
of these matrices, which leads to analytical results that are 
not explicitly covariant. Therefore, for the explicit 
calculation of some elements of the ${\cal S}$ matrix 
which involve fermions, 
there could be an improvement if one can express them within a very 
general, covariant, representation independent formula, where 
the spin degrees of freedom clearly appear, and where the 
limit $m\ra 0$ is meaningful and rather straightforward. 
There already exist in the literature some very useful 
formulas (see~\cite{maina,bouchiat,wudka}), 
and the purpose of this paper is to generalize them, keeping in 
mind that we want to respect the criteria described above. 
For instance, the definition of the ``conjugate'' spinor $\bar \psi$ by 
$\psi^\dagger\gamma^0$ is purely conventional~\cite{jauch}, 
assuming this way that we take a unitary representation of the 
Dirac matrices. This actually is not imposed by a physical principle 
but is rather a way to normalize the lagrangian. 
These matrices are even not supposed to be hermitian or 
anti-hermitian, and most of the calculations in this paper 
are completely representation independent, except 
for the sake of illustration. 

 It is not claimed however that the formulas presented in this 
paper lead to faster algorithms for the calculation of huge 
Feynman amplitudes, and we must recall the reader that for this 
purpose, there are some quite fast calculation 
techniques~\cite{xzc,vincent} using spinor inner products. However, 
these algorithms are very fast only for massless fermions and 
introducing the case of massive fermions requires much more 
complex calculations. It is possible that in some cases, the formulas 
presented in this paper could be competitive for writing Monte-Carlo 
programs. Our purpose here is mostly to give simple tools for 
calculating analytically some amplitudes that are not very large, 
in such a way that one can possibly see the main physical 
features of the amplitude just by looking at its expression. 
For spinor inner products, one makes  
use of polar coordinates which need to set several geometrical 
conventions, yielding calculations that are not explicitly Lorentz 
covariant. This is why, although these tools exist and have proved 
to be efficient, we have found also useful to present how to 
calculate some amplitudes in a way which respect the symmetries 
(physical or not) of the problem and which includes as few  
conventions as possible.
    
The paper is organized as follows. In the first section, we review  
some generalities about the Dirac equation, focusing especially 
on the subtleties that are very scarcely found in elementary 
textbooks, and which we found useful to gather here. Then 
we show on a simple example of calculation involving neutrinos, how 
one can in general express an amplitude in a representation 
independent and explicitly covariant way. The next part 
is devoted to the mathematical derivation of our general 
formulas. For this purpose, we will 
redemonstrate the well-known formula giving the density 
matrix of a pure spinor state $u\bar u = (p\sla+m)(1+
s\sla\gamma^5)/2$, using some representation independent 
calculations. We then show how to compute Feynman amplitudes 
or its square using these projection operator, focusing especially 
on the possible singularities appearing in the phase space which 
must be taken under consideration for an implementation in a 
Monte-Carlo program. The advantage of the demonstration we used is 
that it shows the uniqueness of the decomposition of the 
density matrix, it exhibits clearly the spin degrees of freedom, 
and especially the transverse degrees of freedom naturally 
emerge for massless fermions. 
   
\section{Generalities about the Dirac Equation}
\label{generalities}

\subsection{From Klein-Gordon to Dirac}

 In this section, we will review some basic things about the 
Dirac equation, and we will also focus on some points that 
are present in the literature, but unfortunately in very few papers.

 First, one must recall that the Dirac equation is obtained by 
looking for a factorization of the  Klein Gordon equation:

\be
[\partial_\mu g^{\mu\nu} \partial_\nu +m^2 ]\Psi =0
\ee
provided we choose the ($+$\,$-$\,$-$\,$-$) metric convention, and  
which can be factored into:

\be
(i\partial\sla +m) (i\partial\sla -m)\Psi 
\ee

where $\partial\sla = \partial_\mu \gamma^\mu$, the $\gamma^\mu$ are  
the Dirac matrices ($4\times4$ complex matrices which must obey 
the anticommutation rules $\{\gamma^\mu,\gamma^\nu\} = 2g^{\mu\nu}$).
 Our choice of a specific sign convention 
for the metric is of importance here. If we choose the 
($-$\,$+$\,$+$\,$+$) signature, we must then replace $m$ 
by $im$ or, to keep the mass term real, $\gamma^\mu$ 
by $i\gamma^\mu$, which is not algebraically equivalent to 
$\gamma^\mu$. This could lead to some confusions when one reads 
the large literature about the differences in the structure of the 
two Clifford algebras corresponding to the two possible sign 
convention for the metric, $(1,3)$ or $(3,1)$   
(see~\cite{pezzaglia} and also some more mathematical 
references~\cite{tucker,okubo,dauns,sternberg}).
However, the resulting physics fortunately does 
not depend on this metric convention, and to show this, we shall 
include the two possible sign conventions by rewriting the 
Klein-Gordon equation in the following way:

\be
[\eta\partial_\mu g^{\mu\nu} \partial_\nu +m^2 ]\Psi =0
\ee

where $\eta = +1$ if the signature is $(1,3)$ and 
$\eta = -1$ if the signature is $(3,1)$. Thus the factorization now 
reads:

\be
(i\omega\partial\sla +m) (i\omega\partial\sla -m)\Psi 
\ee

where $\omega$ is one square root of $\eta$, i.e. up to a sign. 
Then, we can also choose between the two operators 
$(i\omega\partial\sla +m)$ and $(i\omega\partial\sla -m)$ 
to define the Dirac equation which gives another sign ambiguity. 
Therefore, to reabsorb these ambiguities, we define 
the matrices $\tilde \gamma^\mu = \omega\gamma^\mu$. We will 
consequently denote $\tilde v = v_\mu \tilde \gamma^\mu$ where 
$v_\mu$ is any 4-vector. The new anticommutation rules are therefore 
$\{\tilde\gamma^\mu,\tilde\gamma^\nu\} = 2\eta g^{\mu\nu}$.  
That is to say, the $\tilde \gamma^\mu$ matrices are in a 
representation of the $C(1,3)$ Clifford algebra whatever the 
sign convention for the metric is. The study of $C(3,1)$, 
with a completely different 
complex structure becomes therefore irrelevant on the physical 
point of view. We can now set the Dirac equation to be 
$(i\tilde \partial -m)\Psi=0$, and since we have just shown 
that the only relevant signature is $(1,3)$, we will assume 
throughout this paper that this metric is chosen (or 
equivalently that $\gamma^\mu = \tilde\gamma^\mu$). 

Since the $\gamma^\mu$ matrices (or $\tilde\gamma^\mu$) do not 
transform like a four-vector, there must be a specific 
transformation for $\Psi$ to ensure the Lorentz covariance 
of this equation\footnote{We recall the reader that by Lorentz 
group is understood the subgroup of $O(1,3)$ which is connected 
to the identity. It is also called the orthochronous 
Lorentz group and denoted $L_+^{\uparrow}$. Spinors are 
in a representation of its covering group $SL(2,C)$ 
(see~\cite{pct,bogoliubov}), but are not in general in a representation 
of the full Lorentz group, covered by what one call the 
$Pin(1,3)$ group~\cite{cdewitt} (and $Pin(3,1)$ for $O(3,1)$). 
The fact that one consider only $L_+^{\uparrow}$ comes from 
the possibility of parity or time reversal violations 
(see~\cite{pct,sachs,tomnio,www,wolfenstein,aharonov.susskind} 
and related experiments~\cite{rauch.al,werner.al}), but 
some authors argue that, for systems that are symmetric under 
the full Lorentz group, one can in principle see experimentally 
some differences between $Pin(1,3)$ and $Pin(3,1)$~\cite{cdewitt}.}:

\be
\Psi(x) \stackrel{\Lambda}{\longrightarrow} 
S_\Lambda\Psi(\Lambda^{(-1)}x) 
=\exp\left(i\frac{\omega_{\mu\nu}}{2}
\sigma^{\mu\nu}\right)\Psi((\Lambda^{(-1)})^\mu_{~\nu}x^\nu)
\ee 

with $\sigma^{\mu\nu} = \frac{i}{2}[\gamma^\mu,\gamma^\nu]$ (or 
equivalently $\frac{i}{2}[\tilde\gamma^\mu,\tilde\gamma^\nu]$), and 
$\omega_{\mu,\nu}$ is an antisymmetric tensor, depending on the 
Lorentz transformation one performs\footnote{$S_\Lambda$ obeys 
$S_\Lambda^{(-1)}\gamma^\mu S_\Lambda = \Lambda^\mu_{~\nu}\gamma^\nu$. 
Using the expansion $e^A B e^{-A} = B + [A,B] +\frac{1}{2!}[A,[A,B]] 
+ \ldots$ and $[\sigma^{\mu,\nu},\gamma^\rho] = 
2i(g^{\nu\rho}\gamma^\mu - g^{\mu\rho}\gamma^\nu)$,  
one gets this relation: 
$\Lambda^\mu_{~\nu} = \exp[-2(\omega)]^\mu_{~\nu}$. }. Thanks to 
this transformation, the Dirac equation 
($(i\tilde\partial -m)\Psi =0$) is Lorentz invariant. 

Now one can question how arbitrary are the Dirac matrices. 
 First, it is easy to check that if we have a set of 4 Dirac 
matrices $\gamma^\mu$, and $S$ is an invertible $4\times 4$ matrix, 
then the new set ${\gamma'}^\mu = S\gamma^\mu S^{-1}$ still 
obey the anticommutation rules, and can be used as well as 
the first set of matrices, provided we operate the 
transformation $\Psi' = S\Psi$ on the wavefunction. The very 
interesting property is that the converse is true, and 
we have summarized in the appendix the old proof given 
in~\cite{jauch}. This property will be used in the following section, 
where we will define the notion of charge conjugation. For this 
purpose, we will also need the general solution of the 
Dirac equation, expressed in momentum space: 
 
\be
\psi = \int \frac{d^3 \vec p}{(2\pi)^3 2p^0} 
\left( a(p)u(p)e^{-ip\cdot x} + b^\dagger(p)v(p)e^{ip\cdot x} \right)
\ee

 where we have included the creation and annihilation operators 
($a(p)$ annihilates a fermion with momentum $p$ and $b^\dagger$ 
creates an antifermion). $u$ and $v$ are the associated spinors that 
are respectively solutions of $(p\sla -m)u(p)=0$ and 
$(p\sla +m)v(p)=0$\footnote{We can remark that we could have chosen 
to write the Dirac equation with a $+m$ instead of $-m$, 
which is equivalent to make the transformation 
$\psi \rightarrow \gamma^5\psi$ on the spinor, and it is in fact 
just a change in the representation of the gamma 
matrices (because it is equivalent to make the transformation 
$\gamma^\mu \rightarrow -\gamma^\mu = \gamma^5 \gamma^\mu \gamma^5 
= \gamma^5 \gamma^\mu (\gamma^5)^{(-1)}$.}. 

\subsection{Spinors and Charge Conjugation}
\label{conjugation}

The Dirac equation being given, we must now construct the 
corresponding lagrangian. For this purpose, we have to face 
the definition of adjoint spinors. Indeed, when one calculate a 
specific Feynman amplitude, one commonly use the definition 
$\bar \psi = \psi^\dagger \gamma^0$ and the property 
${\gamma^\mu}^\dagger = \gamma^0\gamma^\mu\gamma^0$. One must 
be careful to the fact that the use of $\gamma^0$ here 
is only conventional, and has very little to do with physics. 

  The main point is to see that whenever 
$\{\gamma^\mu, \mu \in [\![0..3]\!]\}$ obey the anticommutation 
rules, then $\{{\gamma^\mu}^\dagger, \mu \in [\![0..3]\!]\}$ 
verify also the same anticommutation relations. We have then 
demonstrated (see in appendix) that these two basis are related 
by an interior automorphism, i.e. there exists an invertible 
matrix $H_+$ such that: 

\be
\forall \mu \in [\![0..3]\!]~~{\gamma^\mu}^\dagger = H_+\gamma^\mu 
H_+^{-1}
\ee 

and also it implies (using the Schur lemma) that $H_+$ verify the 
property $H_+^\dagger = {\bf h_+} H_+$ with ${\bf h_+}$ a 
complex number of modulus one. Then the adjoint spinor is defined as 
$\bar \psi = \psi^\dagger H_+$. The Dirac lagrangian can 
then be constructed as follows: 

\be
{\cal L} = \bar \psi (i\partial\sla -m) \psi 
\ee

For the sake of unitarity of the scattering matrix, we need the   
classical lagrangian to be real, which implies ${\bf h_+}=1$, 
that is to say $H_+$ is hermitian\footnote{When one changes 
the representation of the Dirac matrices through $\gamma^\mu = 
S{\gamma'}^\mu S^{-1}$, $H_+$ changes into 
${H'}_+ = \lambda S^\dagger H_+ S$ like an 
hermitian form (its signature is $(2,2)$).}. 
For the class of unitary 
representations, we can take $H_+ = \gamma^0$. However, unitary 
representations have no more physical relevance than 
any other representation, and therefore we shall not suppose in the 
following that $H_+$ is equal to $\gamma^0$. Yet, since 
$H_+$ is defined up to a complex constant, imposing the unitarity 
of $H_+$ is a good way to normalize the lagrangian. We will 
rather impose a normalization on the density matrices at the 
end of section~\ref{solutions}. Although it leads to the 
same constraint at the end, it allows us to keep $H_+$ undetermined. 
In other words, we will normalize the density matrices 
instead of $H_+$.   
 
Now, the notion of charge conjugation needs the introduction of the 
gauge field and we get $(i\partial\sla -eA\sla -m)\psi = 0$ for 
the equation of motion of the interacting fermion field. 
Taking now the complex conjugate we obtain the charge conjugated 
equation $(i\partial\sla +eA\sla -m)\psi^c=0$, provided 
we take the following definition for $\psi^c$:

\bea
\psi^c &=& C_-^{-1}\psi^* \\
\forall \mu \in [\![0..3]\!]~~{\gamma_\mu^*} &=& 
-C_-\gamma_\mu C_-^{-1}
\label{defC}
\ena 

(Note that when one conjugates $\psi$, one shall not forget to 
take the hermitian conjugate of the creation and annihilation 
operators). This definition can be seen at the level of free 
spinors in the momentum representation: if we have $(p\sla -m)u(p)=0$, 
then after complex conjugation 
one gets $(p\sla + m)C_-^{-1}u^*(p)=0$. In other words, the charge 
conjugated spinor of a so-called {\bf u} spinor becomes a {\bf v} 
spinor, which is the intuitive definition of charge conjugation 
(by computing $\psi^c$ one can make the following 
identifications: $u^c = C_-^{-1}v^*$ and $v^c = C_-^{-1}u^*$). 
The existence of $C_-$ is actually given
 by the theorem we have proved in appendix relating sets of matrices 
that obey the anticommutation relations, which is the case of 
$-\gamma_\mu^*$\footnote{In the so-called Majorana representation of 
the Dirac matrices, $C_-$ is just a multiple of the identity matrix, 
and if we look at another representation 
$\gamma'_\mu = S\gamma_\mu S^{-1}$ then the matrix $C_-$ 
transforms with the formula $C_-' = \alpha S^* C_- S^{-1}$, 
where $\alpha$ can be any complex number 
different from $0$.}.
$C_-$ must satisfy $C_-^* C_- = {\bf c_-}I$ where ${\bf c_-}$ is a 
real constant\footnote{In fact, one can scale the matrix $C_-$ such 
that ${\bf c_-} =\pm 1$. The sign of ${\bf c_-}$ depends on 
the convention we choose 
for the metric, but not on the representation of the Dirac matrices. 
Actually, one has ${\bf c_-} =+1$ for the convention 
($+$\,$-$\,$-$\,$-$) (see~\cite{tucker}) and ${\bf c_-} =-1$ if 
one uses the ($-$\,$+$\,$+$\,$+$) convention, 
like in the book of Jauch and Rohrlich~\cite{jauch}. 
In this latter case, charge conjugation becomes an anti-involution, 
and we could wrongly state that Majorana spinors (spinors which 
are identical to their charge-conjugated version) do not exist 
in this case, and the physics would depend on the convention for the 
metric. The point is in this case that the rule given to define  
charge conjugation would change, and we should have with 
this convention $\psi^c = C_+^{-1}\psi$ where 
$\gamma_\mu^* = C_+\gamma_\mu C_+^{-1}$ and $C_+^* C_+ = 
{\bf c_+}I$ (${\bf c_+}>0$). One can obviously relate $C_+$ and 
$C_-$ by $C_+ = C_-\gamma^5$.}. 

Having defined charge conjugation, we can take the opportunity 
to define a Majorana spinor. A Majorana spinor 
is its own charge conjugated particle, which necessarily 
constrains this kind of particle to be neutral, and 
it must therefore obey the relation $\psi^c = \alpha \psi$ 
($\alpha$ being a possible phase). It implies the following 
relations between creation-annihilation operators and {\bf u} 
and {\bf v} spinors:

\bea
b&=&a \label{majorana0} \\
C_-^{-1}{\bf v}^* &=& \alpha {\bf u} \label{majorana1} \\
C_-^{-1}{\bf u}^* &=& \alpha {\bf v} \label{majorana2}
\ena

 And for the sake of consistency, one should have 
$C_-^*C_-=\frac{1}{|\alpha|^2}I$. This happens to be possible because  
the sign of ${\bf c_-}$ is independent of the representation 
of the Dirac matrices, and in the Majorana 
representation of the Dirac matrices\footnote{Note that the Majorana 
representation is only a kind of ``eigenvector'' for complex 
conjugation (up to a real inner automorphism), which is  
different from the notion of Majorana spinors. The existence of a 
Majorana representation (which depends only on the metric signature), 
only implies the possible existence of Majorana spinors.}, $C_-$ 
is proportional to the identity (and thus $C_-$ can always be 
written in the form $C_-= \sqrt{{\bf c_-}} S^*S^{-1}$ where $S$ is any 
$4\times 4$ invertible matrix). 
In fact, eq.~\ref{majorana1} and~\ref{majorana2} 
don't tell us much more about the structure of the wave function 
of a Majorana fermion, and Majorana spinors are nothing but 
standard spinors. The main difference comes from the quantization 
of the field, and especially from eq.~\ref{majorana0} which 
allows to construct specific lagrangians that may distinguish 
between Majorana and Dirac spinors. We won't give more details 
about Majorana fermions here and refer the reader to the 
literature on this topic.

In a similar way, we could also have defined an invertible matrix 
$T_+$ or $T_-$ for the transposition of the Dirac matrices 
(which also obey the anticommutation rules), but the three operations 
are related and since $T_-$ can be obtained by combining  
$H_+$ and $C_-$, we will not have to use it. However, it can be useful 
to get a relation between $C_-$ and $H_+$ using the fact that 
hermiticity and complex conjugation are two commuting operations. 
For this purpose we write:

\bea
{^t}\gamma_\mu = (\gamma_\mu^\dagger)^* &=& -H_+^* C_-\gamma_\mu 
C_-^{-1}(H_+^*)^{-1}~~~~(= -T_-\gamma_\mu T_-^{-1}) \\ 
{^t}\gamma_\mu = (\gamma_\mu^*)^\dagger &=& -(C_-^{-1})^\dagger H_+ 
\gamma_\mu H_+^{-1}C_-^\dagger  
\ena

And using the Schur lemma, we know that there exists a non-vanishing 
constant $\kappa$ such that:

\bea
H_+^*C_- &=& \kappa (C_-^{-1})^\dagger H_+ \label{rel1} \\ 
H_+^{-1} C_-^\dagger &=& \kappa C_-^{-1} (H_+^*)^{-1} \label{rel2}
\ena

Then, from a density matrix like $\rho = \psi\bar \psi$ (or more 
generally a rank~1 matrix like $\psi\bar\psi'$),  
we can compute its charge conjugated version $\psi^c \bar \psi^c$,  
and using the relations written in this section we may obtain easily: 

\bea
\rho^c = \psi^c \bar \psi^c &=& C_-^{-1} \rho^* (H_+^*)^{-1} 
(C_-^{-1})^\dagger H_+ \\  
&=& (\kappa)^{-1} C_-^{-1} \rho^* C_- \label{rho-conj}
\ena

The last equation being obtained thanks to eq.~\ref{rel2}, and 
if we require the lagrangian to be invariant under charge 
conjugation we must set $\kappa = 1$. We will see beyond 
(section~\ref{rhoc}) that charge conjugation, given by 
eq.~\ref{rho-conj} can be expressed as a condition on some 
Lorentz tensors which define completely the density matrix $\rho$.

\section{General fermionic amplitudes}

\begin{figure}
\begin{center}
\caption{\label{ffx}{\em Amplitudes involving some fermion pairs}}
\mbox{
\begin{picture}(20000,12000)
\drawline\fermion[3 0](9000,10000)[5000]
\drawarrow[3\ATTIP](\pmidx,\pmidy)
\global\Xone=\pfrontx \global\Yone=\pfronty
\global\advance\Xone by -2500 \put(\Xone,\Yone){$f$, $p$}
\put(0,6000){${\cal M}_{in,~\overline{in}}(p,p') =$}
\global\Xone=\pbackx \global\Yone=\pbacky
\put(\Xone,\Yone){\circle*{3000}}
\drawline\fermion[5 0](\pbackx,\pbacky)[5000]
\global\Xtwo=\pbackx \global\Ytwo=\pbacky
\global\advance\Xtwo by -2500 \put(\Xtwo,\Ytwo){$\bar f$, $p'$}
\drawarrow[1\ATTIP](\pmidx,\pmidy)
\put(\Xone,\Yone){\line(1,1){4000}}
\put(\Xone,\Yone){\line(2,1){5000}}
\put(\Xone,\Yone){\line(2,0){5500}}
\put(\Xone,\Yone){\line(2,-1){5000}}
\put(\Xone,\Yone){\line(1,-1){4000}}
\end{picture}
\begin{picture}(20000,12000)
\drawline\fermion[3 0](9000,10000)[5000]
\drawarrow[3\ATTIP](\pmidx,\pmidy)
\global\Xone=\pfrontx \global\Yone=\pfronty
\global\advance\Xone by -2500 \put(\Xone,\Yone){$f$, $p$}
\put(0,6000){${\cal M}_{in,~out}(p,p') =$}
\global\Xone=\pbackx \global\Yone=\pbacky
\put(\Xone,\Yone){\circle*{3000}}
\drawline\fermion[5 0](\Xone,\Yone)[5000]
\drawline\fermion[1 0](\Xone,\Yone)[5000]
\global\Xtwo=\pbackx \global\Ytwo=\pbacky
\global\advance\Xtwo by 1000 \global\advance\Ytwo by 500
\put(\Xtwo,\Ytwo){$f$, $p'$}
\drawarrow[1\ATBASE](\pmidx,\pmidy)
\put(\Xone,\Yone){\line(2,1){5000}}
\put(\Xone,\Yone){\line(2,0){5500}}
\put(\Xone,\Yone){\line(2,-1){5000}}
\put(\Xone,\Yone){\line(1,-1){4000}}
\end{picture}
} \mbox{
\begin{picture}(20000,12000)
\drawline\fermion[3 0](9000,10000)[5000]
\drawarrow[3\ATTIP](\pmidx,\pmidy)
\global\Xone=\pfrontx \global\Yone=\pfronty
\global\advance\Xone by -2500 \put(\Xone,\Yone){$\bar f$, $p'$}
\put(0,6000){${\cal M}_{\overline{in},~\overline{out}}(p,p') =$}
\global\Xone=\pbackx \global\Yone=\pbacky
\put(\Xone,\Yone){\circle*{3000}}
\drawline\fermion[5 0](\Xone,\Yone)[5000]
\drawline\fermion[1 0](\Xone,\Yone)[5000]
\global\Xtwo=\pbackx \global\Ytwo=\pbacky
\global\advance\Xtwo by 1000 \global\advance\Ytwo by 500
\put(\Xtwo,\Ytwo){$\bar f$, $p$}
\drawarrow[1\ATBASE](\pmidx,\pmidy)
\put(\Xone,\Yone){\line(2,1){5000}}
\put(\Xone,\Yone){\line(2,0){5500}}
\put(\Xone,\Yone){\line(2,-1){5000}}
\put(\Xone,\Yone){\line(1,-1){4000}}
\end{picture}
\begin{picture}(20000,12000)
\drawline\fermion[3 0](9000,10000)[5000]
\put(0,6000){${\cal M}_{out,~\overline{out}}(p,p') =$}
\global\Xone=\pbackx \global\Yone=\pbacky
\put(\Xone,\Yone){\circle*{3000}}
\drawline\fermion[5 0](\pbackx,\pbacky)[5000]
\drawline\fermion[1 0](\Xone,\Yone)[5000]
\global\Xtwo=\pbackx \global\Ytwo=\pbacky
\global\advance\Xtwo by 1000 \global\advance\Ytwo by 500
\put(\Xtwo,\Ytwo){$f$, $p'$}
\drawarrow[1\ATBASE](\pmidx,\pmidy)
\drawline\fermion[3 0](\Xone,\Yone)[5000]
\global\Xtwo=\pbackx \global\Ytwo=\pbacky
\global\advance\Xtwo by 1000 \global\advance\Ytwo by -500
\put(\Xtwo,\Ytwo){$\bar f$, $p$}
\drawarrow[3\ATBASE](\pmidx,\pmidy)
\put(\Xone,\Yone){\line(1,1){4000}}
\put(\Xone,\Yone){\line(2,1){5000}}
\put(\Xone,\Yone){\line(2,0){5500}}
\put(\Xone,\Yone){\line(2,-1){5000}}
\end{picture} 
}
\end{center}
\end{figure}
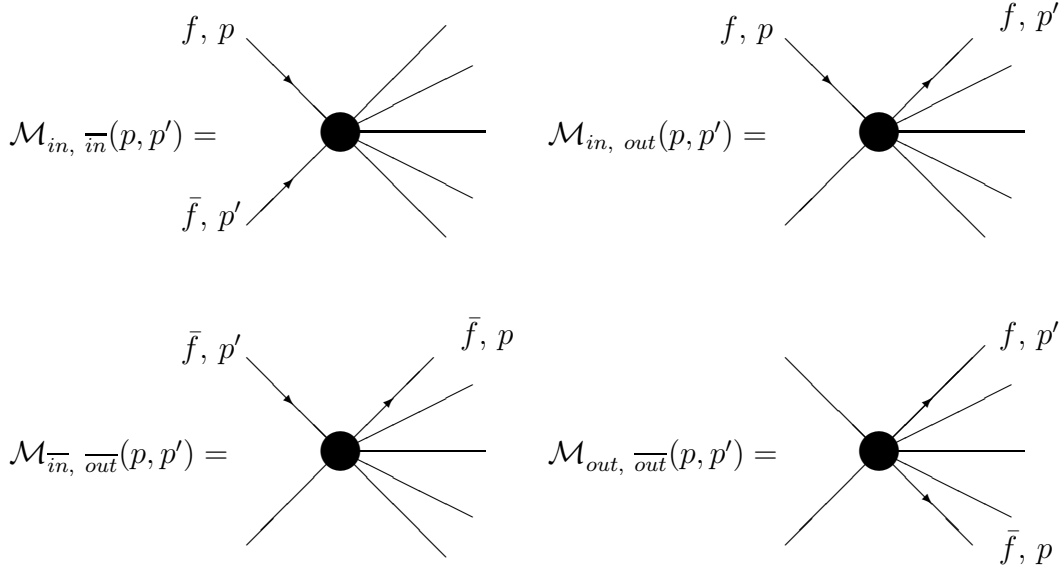

 In this section, we sketch the way we intend to compute 
any scattering amplitude with at least one fermionic current. 
 We shall first note that there are four different possible 
configurations corresponding to {\it a priori} four different 
expressions for the amplitudes (see fig.~\ref{ffx}, where 
the overlined letters correspond to the antiparticles).
 
  In a condensed notation, one may note a general 
amplitude as:

\be
{\cal M} = \bar \psi' O \psi = tr[{\cal O}\psi \bar \psi'] 
= tr[{\cal O}\Sigma]
\ee

With $\Sigma = u(-cp)\bar u(-c'p')$ where $u(p)$ is a 
solution of $(p\sla-m)u=0$, and $c$, $c'$ are the 
fermionic charge (\ie $-1$ for a fermion, $+1$ for 
an antifermion), provided that $p$ is the 
momentum corresponding to an {\bf incoming} fermionic current, 
and $p'$ is the momentum corresponding to an {\bf outgoing} 
fermionic current (see fig~\ref{ffx}). 
Of course, we could have chosen to write an amplitude by reference 
to an antifermionic current, or an annihilation or a 
pair production, but the four type of amplitudes are related 
and it is just a matter of convention. Also, we will  
parameterize the generic amplitudes using $u(-cp)$ for the 
two possible choices (fermions or antifermions) instead of 
using charge conjugation because we will see beyond that 
charge conjugation can be easily expressed in terms of 
changing the sign of the constant $c$ or $c'$ and also 
the sign of the spin degree of freedom which will be introduced   
in the next section. The general expression of the amplitude 
can be specialized to the four kinematic configurations 
of figure~\ref{ffx} in this way:

\bea
{\cal M}_{in,~\overline{in}} &=& {\bar v}(p_+^i){\cal O}
u(p_-^i) = 
tr[{\cal O}u(p_-^i){\bar u}(-p_+^i)]\\
{\cal M}_{in,~out} &=& {\bar u}(p_-^o){\cal O}u(p_-^i) = 
tr[{\cal O}u(p_-^i){\bar u}(p_-^o)]\\
{\cal M}_{\overline{in},~\overline{out}} &=& 
{\bar v}(p_+^o){\cal O}v(p_+^i) 
= tr[{\cal O}u(-p_+^i){\bar u}(-p_+^o)] \\
{\cal M}_{out,~\overline{out}} &=& {\bar u}(p_-^o)
{\cal O}v(p_+^o) 
= tr[{\cal O}u(-p_+^o){\bar u}(p_-^o)]
\ena 

where $p_-$ denotes the momentum of the fermion, 
$p_+$ is for the antifermion, and 
${\cal O}$ is the interaction operator. 
Instead of decomposing spinors in  their components, we  
will express the amplitudes without having to 
specify a specific representation for the Dirac 
matrices\footnote{Some variants of this formulation exist 
already in the literature, see~\cite{maina,bouchiat,wudka}. Another 
calculation technique was developed by Hagiwara and 
Zeppenfeld~\cite{algo.hagiwara}, decomposing spinors  
in the chiral representation into their Weyl spinors. This 
method yields a quick algorithm for computation but here, we are 
looking for a formulation that is explicitly covariant 
and representation independent.}. 
We shall therefore try to express the rank~1 matrix 
$\Sigma = u(-cp)\bar u(-c'p')$ in the basis of Dirac matrices, and 
 we can do the same for the operator ${\cal O}$. Then the 
trace can be easily computed using simple algorithms\footnote{Symbolic 
calculation programs can do this very easily nowadays. The fastest 
being probably FORM, and a rather convenient one is  
``M''\cite{M}.}. If one prefer to compute directly the square 
of the amplitude, one has: 

\bea
|{\cal M}|^2 &=& \left[\bar u(-c'p') {\cal O} u(-cp)\right]
\left[\bar u(-c'p') {\cal O} u(-cp)\right]^\dagger \nonumber \\
&=& \left[\bar u(-c'p') {\cal O} u(-cp)\right] 
\left[\bar u(-cp)H_+^{-1}{\cal O}^\dagger H_+ u(-c'p')\right] 
\nonumber \\ 
 &=& tr\left[ \rho'{\cal O}\rho \bar{\cal O} \right] \label{ampl-sq}
\ena

with $\rho = u(-cp)\bar u(-cp)$, $\rho' = u(-c'p')\bar u(-c'p')$ 
and $\bar {\cal O} = H_+^{-1}{\cal O}^\dagger H_+$. If we decompose 
${\cal O}$ on the basis of the Dirac matrices we have: 

\be
\label{o-decomp}
{\cal O} = s_o+\bar s_o\gamma^5+{v\sla}_o
+{a\sla}_o\gamma^5 + S^o_{\mu\nu}\sigma^{\mu\nu}
\ee

and\footnote{By ${v\sla}^*$ we denote $v^*_\mu \gamma^\mu$. The Dirac 
matrices are not conjugated in this notation.}: 

\be
\label{obar-decomp}
\bar{\cal O} = s_o^*-\bar s_o^*\gamma^5+{v\sla}^*_o
+{a\sla}^*_o\gamma^5 + {S^o_{\mu\nu}}^*\sigma^{\mu\nu}
\ee

 Then, in the following sections we will give the decomposition 
on the same basis for the density matrix $\rho$, 
which will give the final result for $\Sigma$ and ${\cal M}$. 
Before, we shall discuss a simple example to show the practical 
procedure we will use in the general case. 

\section{A simple example}
\label{example} 

 We will thus consider in this section the scattering of a neutrino, 
for instance in an electroweak process. The neutrino spinors can 
be easily expressed in terms of Weyl spinors if we take the chiral 
representation for the Dirac matrices. We get:

\be
\Phi_L=\left(\begin{array}{c} -e^{-i\varphi/2}\sin(\theta/2)\\
e^{i\varphi/2}\cos(\theta/2) \end{array}\right),~\Phi_R=
\left(\begin{array}{c} e^{-i\varphi/2}\cos(\theta/2)\\
e^{i\varphi/2}\sin(\theta/2) \end{array}\right)
\ee

and using the notations $p_{(in,out)} = 
E_{(in,out)}(1, \vec n_{(in,out)})$, the $\Sigma$ matrix can 
be computed with a little bit of algebra:

\bea
u(p_{in})\bar u(p_{out})&=&\sqrt{E_{in}E_{out}}\left(\begin{array}{cc} 
0 &0 \\ \Phi_L(p_{in})\Phi^\dagger_L(p_{out})
=\Omega & 0 \end{array}\right) \\
 &=& \sqrt{E_{in}E_{out}}\gamma^0\left(\begin{array}{cc} 
\Omega &0 \\ 0 & 0 \end{array}\right)
= \ldots = \frac{1-\gamma^5}{2}v\sla 
\ena

One may then define the vector $v^\mu$ such 
that $v^0=(\Omega_{11}+\Omega_{22})/2$, 
$v^1=-(\Omega_{12}+\Omega_{21})/2$, 
$v^2=-i(\Omega_{12}-\Omega_{21})/2$, 
$v^3=-(\Omega_{11}-\Omega_{22})/2$. When the two fermions are 
not back-to-back, $v^0$ doesn't vanish and one can impose 
the condition that $v^0$ is real, multiplying by the 
suitable phase. One finally obtains the following form 
for $v^\mu$:

\be
\label{v-exp}
v^\mu =\sqrt{E_{in}E_{out}}\left(\begin{array}{c} 
\sqrt{2}\sqrt{1+\vec n_{in}\cdot\vec n_{out}}\\
\ds -\frac{\sqrt{2}}{\sqrt{1+\vec n_{in}\cdot\vec n_{out}}}
(\vec n_{in}+\vec n_{out} 
+i\vec n_{in}\wedge\vec n_{out}) \end{array}\right)
\ee
 
One can see that $v^\mu$ is orthogonal to both momenta 
of the fermions, and moreover, that the space-like part 
of this vector represents an elliptic polarization 
associated to the 3-vector $\vec n_{in}-\vec n_{out}$.
We have therefore combined the two spin~$(1/2)$ particles to obtain 
a spin-1 current.

  But we will show that the expression of $v^\mu$ can be 
expressed in a covariant way through the formula:

\be
v^\mu =\frac{(p_{out}\cdot k)p_{in}^\mu + (p_{in}\cdot k)p_{out}^\mu
-(p_{out}\cdot p_{in})k^\mu -i\eps^{\mu\nu\rho\sigma}
k_\nu p_{in\,\rho} p_{out\,\sigma}}{\sqrt{2}\sqrt{2(p_{out}\cdot k)
(p_{in}\cdot k)-(p_{out}\cdot p_{in})k^2}}
\ee

Where the four-vector $k^\mu$ serves as a reference. 
Expression~\ref{v-exp} is actually obtained by setting 
$k^\mu = (1,\vec 0)$, but one can note that $v^\mu$ is 
invariant under the transformation 
$k~\ra~k+\lambda\,p_{in}~+\mu\,p_{out}$, which also implies 
that $k$ must not lie in the plane generated by $p_{in}$ and 
$p_{out}$. This will be rather obvious with the following. Note 
also that a rescaling of $k$ does not affect $v$. Therefore, there 
is only one degree of freedom of $k$ that may influence $v$, 
through a phase multiplication. 

 To obtain this expression, let's consider the density  
matrix for both the incoming and outgoing momenta (we will 
derive its expression in the following section): 
\be
 u {\bar u} = {p\sla}\frac{1+\gamma^5}{2}
\ee 

then one has for a given vector $k^\mu$:

\be
\label{density-prod}
u_{in} {\bar u}_{in}  k\sla  u_{out} {\bar u}_{out} 
= ({\bar u}_{in} k\sla  u_{out}) \Sigma = 
{p\sla}_{in} k\sla{p\sla}_{out}\frac{1+\gamma^5}{2} = 
v\sla\frac{1+\gamma^5}{2}
\ee

with: 

\be
v^\mu = (p_{out}\cdot k)p_{in}^\mu + (p_{in}\cdot k)p_{out}^\mu
-(p_{out}\cdot p_{in})k^\mu -i\eps^{\mu\nu\rho\sigma}
k_\nu p_{in\,\rho} p_{out\,\sigma}
\ee

If $k$ lies in the plane generated by $p_{in}$ and $p_{out}$,  
$v$ vanishes and the latter relations are useless. So let's 
take $k$ outside this plane. Then we obtain the normalization 
factor in eq.~\ref{density-prod} by multiplying once again 
by ${k\sla}^*$\footnote{${k\sla}^* \stackrel{def}{=} 
k^{*\mu}\gamma_\mu$ (see previous section).} and taking the trace, 
we get:

\bea
|{\bar u}_{in} k\sla  u_{out}|^2
 &=& ({\bar u}_{in} k\sla u_{out}) ({\bar u}_{out} {k\sla}^*  u_{in}) 
 \nonumber \\
&=& tr\left[v\sla\frac{1+\gamma^5}{2}{k\sla}^*\right] = 2(v\cdot k^*)
\nonumber \\
&=& 2\left(2\Re((p_{in}\cdot k)(p_{out}\cdot k^*)) 
-(p_{out}\cdot p_{in})(k\cdot k^*)\right. \nonumber \\
&& \left. +i\eps^{\mu\nu\rho\sigma} 
k_\mu k^{*}_\nu p_{in\,\rho} p_{out\,\sigma}\right) 
\ena
\bea
\lefteqn{\Rightarrow {\bar u}_{in} k\sla  u_{out} =} \nonumber \\ 
&& e^{i\varphi} \sqrt{2}\sqrt{2\Re((p_{in}\cdot k)(p_{out}\cdot k^*)) 
-(p_{out}\cdot p_{in})(k\cdot k^*) +i\eps^{\mu\nu\rho\sigma}
k_\mu k^{*}_\nu p_{in\,\rho} p_{out\,\sigma}} 
\ena

In this example, we have been obliged to introduce another 
4-vector which has {\it a priori} no physical meaning (its role is 
only to set a reference for the phase of the amplitude), and we 
can be led into trouble if this 4-vector is not suitably chosen. 
This problem would not occur if we had computed directly the square 
of the amplitude, for which the introduction of this arbitrary 
4-vector wouldn't be necessary. This also enlighten the impossibility  
to find a formalism for the amplitude which would be continuous in the 
massless limit, since this kind of singularity does not exist 
if one of the fermions is massive and $k$ is chosen on the light 
cone. We will discuss this point further 
at the end of the paper, and we now turn to the general case 
of an amplitude that includes a fermionic current, represented 
by what we have called the $\Sigma$ matrix.

\section{The general case}

\subsection{General properties of the $\Sigma$ matrix}

  In this section, we will specify the properties of 
the $\Sigma$ matrix. Instead of expressing the spinors 
in a specific representation of the the Dirac matrices, 
we will express the general equations that determine 
$\Sigma$ without having to take any specific representation.
 These equations are therefore more physically meaningful. 
This kind of formalism have already been studied a very long time 
ago by Pauli and others. It was called the bispinor algebra, 
and their intrinsic relations were studied in the context 
of Fierz identities (see~\cite{crawford}). 
 We just give here another demonstration, which does not use 
Fierz identities which contain far too much information 
for our purpose. The main point of our proof being in fact a 
representation independent characterization 
of a rank-1 $4\times 4$ matrix. To proceed in the same way 
as in the previous example, we shall deal first with 
the density matrices $\rho$, for which we will use three 
constraints. The first is the equation of motion 
$(p\sla-m)\rho = 0 = \rho(p\sla-m)$. Then we will 
use a conjugation constraint $\bar \rho\, \stackrel{def}{=} \,
H_+^{-1}\rho^\dagger H_+ = \rho$, and finally 
 a criterion to say that $\rho$ is a rank~1 matrix (using 
a characterization described in appendix). In principle, 
after these three steps, there should only remain a normalization 
condition to determine. So, let us first write the consequences 
of the equations of motion on $\rho$, provided we have 
decomposed $\rho$ on a typical basis of Dirac matrices, that is: 

\be
\label{deux}
\rho = s+\bar s\gamma^5+{v\sla}+{a\sla}\gamma^5 +
S_{\mu\nu}\sigma^{\mu\nu}
\ee

where $S_{\mu\nu}$ is an antisymmetric tensor, 
$\sigma^{\mu\nu}=\frac{i}{2}[\gamma^\mu,\gamma^\nu]$, 
 and one has the relations: 

$\begin{array}{cccccc}
\\
s &=&\frac{1}{4}tr[\rho] & \bar s &= &  
\frac{1}{4}tr[\rho\gamma^5] \\
\\
v^\mu & = & \frac{1}{4}tr[\rho\gamma^\mu] & a^\mu & = & 
 \frac{1}{4}tr[\rho\gamma^5\gamma^\mu] \\
\\
S_{\mu\nu} & = & \frac{1}{8}tr[\rho\sigma^{\mu\nu}] & & & \\
\end{array}$

Then the equations of motion can be also projected on the same basis 
and it gives \footnote{$[p,v]^{\mu\nu} \stackrel{def}{=} p^\mu v^\nu 
- p^\nu v^\mu$}:

\bea
(p\sla-m)\rho &=& 0 \Leftrightarrow \\
p\cdot v &=& m s \\
p\cdot a &=& m\bar s \\
sp^\mu -2iS^{\mu\nu}p_\nu &=& mv^\mu \\
\bar s p^\mu -\eps^{\alpha\beta\mu\nu}S_{\alpha\beta}p_\nu &=& 
ma^\mu \\ 
\frac{1}{2}\eps^{\mu\nu\rho\sigma}a_\rho p_\sigma 
-\frac{i}{2}[p,v]^{\mu\nu} &=& mS^{\mu\nu} 
\ena

and: 

\bea
\rho(p\sla-m) &=& 0 \Leftrightarrow \\
p\cdot v &=& m s \\
p\cdot a &=& -m\bar s \\
sp^\mu +2iS^{\mu\nu}p_\nu &=& mv^\mu \\
-\bar s p^\mu -\eps^{\alpha\beta\mu\nu}S_{\alpha\beta}p_\nu &=& 
ma^\mu \\ 
\frac{1}{2}\eps^{\mu\nu\rho\sigma}a_\rho p_\sigma 
+\frac{i}{2}[p,v]^{\mu\nu} &=& mS^{\mu\nu} 
\ena

These two sets of equations can be rearranged into a simpler one:

\bea
\bar s &=& 0 \label{simple1} \\ 
\ [p,v]^{\mu\nu} &=& 0 \Leftrightarrow v^\mu = \lambda p^\mu 
\label{simple2} \\ 
sp^\mu &=& mv^\mu \Rightarrow s = \lambda m \label{simple3} \\ 
p\cdot a &=& 0 \label{simple4} \\ 
p_\mu S^{\mu\nu} &=& 0  \label{simple5} \\ 
\eps^{\mu\nu\alpha\beta}S_{\alpha\beta}p_\nu &=& -ma^\mu 
\label{simple6} \\
\eps^{\mu\nu\alpha\beta} a_\alpha p_\beta &=& 2m S^{\mu\nu} 
\label{simple7}
\ena

In the massive case, the two last equations are indeed equivalent. 
We can already see that the structure of the density matrix is 
quite constrained. Since $\bar s =0$, the conjugation constraint 
simply says that all the remaining coefficients ($s$, $v^\mu$, 
$a^\mu$ and $S_{\mu\nu}$) must be real (see eq.~\ref{obar-decomp}). 
We shall now use the 
characterization of $\rho$ as a rank~1 matrix, which is the last 
constraint on $\rho$. In order to find the necessary and sufficient 
conditions for $\rho$ to be a rank~1 matrix, we will apply the 
corollary of the second theorem demonstrated in appendix~B, and 
in the first place we shall not make reference to the constraints 
we have already obtained with the equations of motion. We will add 
them at the end. The corollary tells us that we must have 
$tr(\rho Q\rho Q') = tr(\rho Q)tr(\rho Q')$ 
for every matrix $Q$ and $Q'$. We will therefore choose for 
$Q$ and $Q'$ the different kind of Dirac matrices, which leads 
to\footnote{We use here the convention $\eps_{0123}=1$.}:

$\star$ For $Q=I$ we obtain respectively for 
$Q'=I,~\gamma^5,~\gamma^\mu, ~\gamma^\mu\gamma^5,~\sigma^{\mu\nu}$:
\bea
\label{un-1}
2S_{\alpha\beta}S^{\alpha\beta} &=& 3s^2-\bar s^2-(v^2-a^2) \\
\label{un-2}
2s\bar s &=& -i\eps^{\mu\nu\rho\sigma}S_{\mu\nu}S_{\rho\sigma} \\
\label{un-3}
sv^\mu &=& \eps^{\lambda\alpha\beta\mu}a_\lambda S_{\alpha\beta} \\
\label{un-4}
sa^\mu &=& \eps^{\lambda\alpha\beta\mu}v_\lambda S_{\alpha\beta} \\  
\label{un-5}
2sS^{\mu\nu} &=& -\eps^{\alpha\beta\mu\nu}(v_\alpha a_\beta + 
i{\bar s}S_{\alpha\beta})
\ena

Note that if $s\ne 0$, the lemma demonstrated in the appendix tells 
us that these 5 conditions~\ref{un-1} to~\ref{un-5} are sufficient 
to ensure that $\Sigma$ is a rank 1 matrix. However, to treat also 
the case where $s=0$, we must set $Q$ to the other kind of ``Dirac 
matrices''. 

$\star$ For $Q=\gamma^5$ we obtain ($Q'=\gamma^5,~\gamma^\mu, 
~\gamma^\mu\gamma^5,~\sigma^{\mu\nu}$):
\bea
\label{deux-2}
2S_{\alpha\beta}S^{\alpha\beta} &=& 3{\bar s}^2-s^2+(v^2-a^2) \\
\label{deux-3}
{\bar s}v^\mu &=& 2i a_\alpha S^{\alpha\mu} \\ 
\label{deux-4}
{\bar s}a^\mu &=& 2i v_\alpha S^{\alpha\mu} \\ 
\label{deux-5}
2{\bar s}S^{\mu\nu} &=&-is\eps^{\mu\nu\rho\sigma}S_{\rho\sigma} 
+i(v^\mu a^\nu-v^\nu a^\mu) \nonumber \\
&& +\frac{1}{4}(S^\mu_{~\alpha}S_{\rho\sigma}\eps^{\nu\alpha\rho\sigma}
 -S^\nu_{~\alpha}S_{\rho\sigma}\eps^{\mu\alpha\rho\sigma})
\ena

$\star$ For $Q=\gamma^\alpha$ and 
$Q'=\gamma^\beta,~\gamma^\beta\gamma^5,~\sigma^{\mu\nu}$ we get:

\bea
\label{trois-3}
0&=&g^{\alpha\beta}(-a^2-v^2+s^2-{\bar s}^2
+2S_{\rho\sigma}S^{\rho\sigma})
-2v^\alpha v^\beta+2a^\alpha a^\beta
+8S^\alpha_{~\lambda}S^{\lambda\beta} \\
\label{trois-4}
0&=& g^{\alpha\beta}(a\cdot v)+(v^\alpha a^\beta -v^\beta a^\alpha)
-s\eps^{\alpha\beta\rho\sigma}S_{\rho\sigma}
+2i{\bar s}S^{\alpha\beta} \\
\label{trois-5}
0&=& i{\bar s}(a^\mu g^{\alpha\nu}-a^\nu g^{\alpha\mu})
+sa_\lambda\eps^{\lambda\alpha\mu\nu} \nonumber \\
&&+2v_\lambda(S^{\lambda\mu}g^{\alpha\nu}-S^{\lambda\nu}g^{\alpha\mu})
+2(-v^\alpha S^{\mu\nu}+v^\mu S^{\alpha\nu}-v^\nu S^{\alpha\mu})
\ena

$\star$ For $Q=\gamma^\alpha\gamma^5$ and $Q'=\gamma^\beta\gamma^5,  
~\sigma^{\mu\nu}$, we obtain:
\bea
\label{quatre-4}
0&=&g^{\alpha\beta}(-a^2-v^2-s^2+{\bar s}^2-2
S_{\rho\sigma}S^{\rho\sigma}) +2v^\alpha v^\beta 
-2a^\alpha a^\beta -8S^\alpha_{~\lambda}S^{\lambda\beta} \\
\label{quatre-5}
0&=& i{\bar s}(g^{\alpha\mu}v^\nu-g^{\alpha\nu}v^\mu) 
-sv_\lambda\eps^{\lambda\alpha\mu\nu} \nonumber \\
&&+2a_\lambda(g^{\alpha\mu}S^{\lambda\nu}-g^{\alpha\nu}S^{\lambda\mu})
+2(a^\alpha S^{\mu\nu} -a^\mu S^{\alpha\nu}+a^\nu S^{\alpha\mu})
\ena

$\star$ For $Q=\sigma^{\mu\nu}$ and $Q'=\sigma^{\rho\sigma}$, 
we obtain:

\bea
\label{cinq-5}
0&=& (g^{\mu\rho}g^{\nu\sigma}-g^{\mu\sigma}g^{\nu\rho})
(s^2+{\bar s}^2+v^2-a^2+2S_{\alpha\beta}S^{\alpha\beta}) \nonumber \\
&&+2\left( g^{\mu\rho}(a^\nu a^\sigma-v^\nu v^\sigma)
 -g^{\mu\sigma}(a^\nu a^\rho-v^\nu v^\rho)
 -g^{\nu\rho}(a^\mu a^\sigma-v^\mu v^\sigma)
 +g^{\nu\sigma}(a^\mu a^\rho-v^\mu v^\rho)  \right) 
\nonumber \\
&& +8( g^{\nu\sigma}S^{\mu}_{~\lambda}S^{\lambda\rho}
     -g^{\nu\rho}S^{\mu}_{~\lambda}S^{\lambda\sigma}
     -g^{\mu\sigma}S^{\nu}_{~\lambda}S^{\lambda\rho}
     +g^{\mu\rho}S^{\nu}_{~\lambda}S^{\lambda\sigma} ) 
\nonumber \\
&& -8(S^{\mu\nu}S^{\rho\sigma}+S^{\mu\sigma}S^{\nu\rho}
-S^{\mu\rho}S^{\nu\sigma} ) -2is{\bar s}\eps^{\mu\nu\rho\sigma}
\ena

In eq.~\ref{trois-5} we can insert eq.~\ref{deux-4} and it 
simplifies into:

\be
v^\alpha S^{\mu\nu}-v^\mu S^{\alpha\nu}+v^\nu S^{\alpha\mu} 
= \frac{s}{2}a_\lambda\eps^{\lambda\alpha\mu\nu}
\ee

And similarly, inserting eq.~\ref{deux-3} into eq.~\ref{quatre-5} 
we get:

\be
a^\alpha S^{\mu\nu}-a^\mu S^{\alpha\nu}+a^\nu S^{\alpha\mu} 
= \frac{s}{2}v_\lambda\eps^{\lambda\alpha\mu\nu}
\ee

Realizing the sum of eq.~\ref{trois-3} and 
eq.~\ref{quatre-4} we obtain $a^2=-v^2$, which then leads to 
$2S_{\alpha\beta}S^{\alpha\beta} = 2(s^2+{\bar s}^2)$ and 
$v^2 = s^2-{\bar s}^2$ using eq.~\ref{un-1} and eq.~\ref{deux-2}. 
The latter relations can be inserted in eq.~\ref{trois-3} and 
eq.~\ref{quatre-4}, the difference of which leads now to the 
simpler relation: 

\be
8S^\alpha_{~\lambda}S^{\lambda\beta} = 
2(v^\alpha v^\beta -a^\alpha a^\beta -s^2 g^{\alpha\beta}) 
\label{eps-normalization}
\ee

Inserting this equation and the previous ones 
in eq.~\ref{cinq-5} we obtain:

\be
 S^{\mu\nu}S^{\rho\sigma}+S^{\mu\sigma}S^{\nu\rho}
-S^{\mu\rho}S^{\nu\sigma} =  -i\frac{s{\bar s}}{4}
 \eps^{\mu\nu\rho\sigma}
\ee

\subsection{Solutions of the rank~1 equations}

If we have the conditions $s\ne 0$ and $s^2-{\bar s}^2\ne 0$, the 
solutions of equations~\ref{un-1} to~\ref{un-5} are given by:

\bea
\label{f1}
a\cdot v &=&0 \\
\label{f2}
v^2 &=& s^2-{\bar s}^2 = -a^2 \\
\label{f3}
S^{\mu\nu} &=& \frac{-i}{2(s^2-{\bar s}^2)}\left[ 
{\bar s}(v^\mu a^\nu -v^\nu a^\mu) -is\eps^{\mu\nu\alpha\beta}
v_\alpha a_\beta \right]
\ena 

The dual of $S^{\mu\nu}$ then reads:

\be
S^{*\mu\nu} = \frac{i}{2}\eps^{\mu\nu\rho\sigma}S_{\rho\sigma}
 = \frac{-i}{2(s^2-{\bar s}^2)}\left[
-s(v^\mu a^\nu -v^\nu a^\mu) +i{\bar s}\eps^{\mu\nu\alpha\beta}
v_\alpha a_\beta \right] 
\ee

 However, the relation~\ref{f1} must always be true thanks to 
equation~\ref{trois-4}. The sum of eq.~\ref{trois-3} and 
eq.~\ref{quatre-4} yields $a^2=-v^2$ in any case, and the 
difference between eq.~\ref{deux-2} and eq.~\ref{un-1} shows 
that the fundamental relation~\ref{f2} must also be true in any case. 
 Note also that in eq.~\ref{deux-5}, the last term could be 
surprising, but using~\ref{f3}, one finds that 
its contribution is $0$. The remaining part of this equation 
is actually the ``dual'' equation of equation~\ref{un-5}.

 Now what happens if ${\bar s} = \pm s$. We can rewrite 
equation~\ref{un-5} using the new notations $S_\pm = S\pm S^*$ 
and $\omega_\pm = s\pm {\bar s}$:

\be
\omega_\pm S_\pm^{\mu\nu} = 
\pm \frac{1}{2}\left[ i(v^\mu a^\nu-v^\nu a^\mu) 
\mp \eps^{\mu\nu\rho\sigma}v_\rho a_\sigma\right]
\ee

 Then if $\omega_+$ or $\omega_-$ vanish, the corresponding 
self-dual tensor is left undetermined by this equation and a 
constraint appear between $v$ and $a$.

\subsection{Solution with the whole set of constraints}
\label{solutions}
 At this stage we can summarize all the constraints we have obtained. 
The equations of motion for the density matrix led us in the massive 
case to:  

\be
\rho = \lambda(m+p\sla)+a\sla\gamma^5 
+ \frac{1}{2m}\eps^{\mu\nu\alpha\beta}a_\alpha p_\beta \sigma_{\mu\nu}
\ee
 
with $a\cdot p=0$ and the rank~1 characterization only adds 
the condition $a^2 = -\lambda^2 m^2$ because of course, 
a lot of the previous equations are redundant. The last two 
equations show that the 4-vectors $a/\lambda \pm p$ 
lie on the light cone. That is to say 
we can write $a = \lambda(p-k_1) = \lambda (k_2-p)$ with 
$k_1^2=k_2^2 =0$, $k_1\cdot p = k_2\cdot p = m^2$. Actually, only 
one of these two vectors is useful since, if $k_1$ is given, 
one can take $2p-k_1$ for $k_2$. Therefore, 
let us consider any non-vanishing 4-vector $k$ lying on the 
light cone. We can therefore set: 

\be
a^\mu = h\lambda\left( p^\mu -\frac{m^2}{p\cdot k}k^\mu\right) 
\label{a-mu}
\ee

Where $h=\pm 1$.   
The scale of $k$ is of no importance in this expression, whereas  
its direction becomes a good candidate for defining the direction 
of reference for the polarization of the fermion. We can also look 
at the light-cone limit of this expression. If the space-like 
projection of $p^\mu$ does not tend to the one of $k^\mu$, the limit 
is obviously the limit of $h\lambda p^\mu$, and the possible 
singularity occurs when $p$ becomes almost collinear to $k$. In order 
to study this limit, we can set $p^\mu = k^\mu + \epsilon \delta^\mu$ 
with $\epsilon \ra 0$ (we must have chosen $k$ suitably here, and 
especially $k^0 >0$). The expression of $a^\mu$ then reads: 

\be
\frac{a^\mu}{h\lambda} = -k^\mu\left(1+\epsilon \frac{\delta^2}
{\delta\cdot k}\right) +\epsilon\delta^\mu
\ee
 
And the limit for $a^\mu$ exists ($a^\mu \simeq -h\lambda p$). 
Rewriting the whole density matrix we get:

\bea
\rho &=& \lambda\left(m+p\sla+h\left(p\sla-\frac{m^2}{p\cdot k}
k\sla\right)\gamma^5 + \frac{mh}{2(p\cdot k)}\eps^{\mu\nu\alpha\beta}
p_\alpha k_\beta \sigma_{\mu\nu}\right) \label{rho} \\  
&=& \lambda\left(m+p\sla+h\left(p\sla-\frac{m^2}{p\cdot k}k\sla\right)
\gamma^5 + mh\left(1-\frac{p\sla k\sla}{p\cdot k}\right)
\gamma^5\right) \\ 
&=& \lambda(p\sla +m)\left(1+ h\left(1-\frac{mk\sla}{p\cdot k}\right)
\gamma^5\right) 
\ena

 We recover for the massive case the well known 
result ($\rho = (p\sla+m)\frac{1+a\sla\gamma^5}{2}$ with 
$a\cdot p =0$ and $a^2 = -1$), in a slightly 
different form. This last formula is already very 
often used in the literature, and one 
may question about the interest in redemonstrating it. 
The first interest is to use this formula (especially in the form 
of eq.~\ref{rho}) to compute scattering amplitudes or the modulus 
square of these amplitudes with external fermions, 
(as we will see in the following) in a way that respects the 
criteria we have defined in introduction. And also, it shows 
naturally how the spin degrees of freedom emerge and the possibility of 
some transverse degrees of freedom in the massless limit. In fact, 
when one looks at the limit $m \ra 0$ in eq.~\ref{rho}, the 
last term vanish\footnote{This limit is obvious if $p\cdot k$ does 
not tend to $0$. Otherwise, if $p\cdot k\ra 0$ when $m\ra 0$, one 
has $p^\mu \ra C k^\mu$. Then one can see that the limit is $0$ 
by taking for instance $p^\mu=(E=\sqrt{m^2+p^2},0,0,p)$ and 
$k = (1,0,0,1)$. Then $\eps^{1203}(p_0k_3-p_3k_0) = E-p = p\cdot k$ 
and the limit is clearly $0$ also in this case.}, and if 
$p\cdot k$ doesn't tend to $0$, 
the term containing $\gamma^5$ becomes equivalent 
to $p\sla\gamma^5$, whereas when 
$p\cdot k$ also tends to $0$ when $m$ vanish (i.e. $p^\mu\ra 
C k^\mu$), then the $\gamma^5$ term is equivalent 
to $-p\sla\gamma^5$. The $m\ra 0$ limit allows us to recover 
the two possible density matrices used in the literature for 
massless fermions ($\rho = p\sla(1\pm\gamma^5)/2$). However, in the 
massive case we had more freedom for this density matrix because the 
direction of $k$ is free. In some sense, this freedom seems 
``frozen'' in the limit $m\ra 0$, but it is not really the case, 
and we shall study in details the solutions of the constraint 
equations for $m=0$ exactly. We can actually deduce from 
eq.~\ref{simple6} that the generic solution for $S^{\mu\nu}$ 
when $m=0$ is $S^{\mu\nu} = \lambda[p,\eps]^{\mu\nu}/2$ with 
$p\cdot \eps =0$ (eq.~\ref{simple5}). 
Therefore, $S^{\mu\nu}$ is not necessarily 
$0$, contrary to what we have deduced from the limit $m\ra 0$. 
The rank~1 conditions then add another constraint, but relax also one. 
Since $a^2=0$ in our case, the normalization of $a^\mu$ is no 
more imposed (the coefficient $\lambda$ in factor of eq.~\ref{a-mu} 
is no more valid) and from eq.~\ref{simple7} one has only 
$a^\mu = \lambda'p^\mu$ with {\it a priori} 
$\lambda'\ne \lambda$. From eq.~\ref{eps-normalization} we 
get a constraint on the normalization of $\eps^\mu$: 

\be
{\lambda}^2(1+\eps^2) = {\lambda'}^2 \label{eps-norm} 
\ee 

We can therefore write the density matrix in the massless limit 
in the form: 

\bea 
\rho &=& p\sla\left(\lambda +\lambda'\gamma^5+i\eps\sla \right) 
\label{massless0} \\
 &=& \lambda p\sla\left( 1+h\sqrt{1+\eps^2}\gamma^5 +i\eps\sla\right) 
\label{massless1}
\ena

The emergence of the ``transverse'' degrees of freedom in the 
massless case is not really a surprise. Their origin comes 
from the fact that the little group of a massless momentum 
($ISO(2)$) is of the same dimension ($3$) as the one of a massive 
momentum ($SO(3)$), and we should therefore obtain the same 
number of degrees of freedom. However, the transverse degrees 
of freedom of massless particles are not observed in experiments, and 
we are led to discard them for neutrinos (see \cite{weinberg-book} 
eq.~2.5.38). In the context of the study of some new physics where 
some massless fermions not yet observed could exist, there is 
{\it a priori} no reason to discard them and it is also a   
reason why we present the most general situation in this paper. 
Finally, for the global normalization of these density matrices, 
we choose $\lambda = 1/2$ to conform to a common normalization 
used in the literature. 

\subsection{Charge conjugation in this formulation}
\label{rhoc}

 Since we have obtained the decomposition of the density matrices 
on the basis of Dirac matrices, we shall now study how charge 
conjugation operates on these decompositions. From what we have 
seen in section~\ref{conjugation} we can write:

\bea
\rho^c &=& C_-^{-1}\rho^*C_- \\ 
&=& s^* I - {\bar s}^*\gamma^5 - {v\sla}^* + {a\sla}^*\gamma^5 
-S_{\mu\nu}^*\sigma^{\mu\nu} \\ 
&=& s I - {v\sla} + {a\sla}\gamma^5 
-S_{\mu\nu}\sigma^{\mu\nu} 
\ena 

The last equation comes from the hermiticity of $\rho$ and the 
fact that we have $\bar s =0$ in any case. Applying this on the 
explicit expression of $\rho$ (eq.~\ref{rho}), the charge 
conjugation acts rather simply on $\rho$: 

\be
\rho(p,k,h) \ra \rho^c(p,k,h) = \rho(p'=-p,k'=k,h'=-h) \label{conj}
\ee

or also this way:

\be
\rho(p,k,h) \ra \rho^c(p,k,h) = 
\rho\left(p'=-p,k'=2p-\frac{m^2}{p\cdot k}k,h'=h\right) 
\ee

We can see that it is equivalent to change the sign of  
 $h$, keeping the vector of reference constant, or 
to keep $h$ the same and to enforce a symmetry operation 
on $k$. Also eq.~\ref{conj} has the advantage to apply in the 
massless case where we can also write from eq.~\ref{massless1}:

\be
\rho(p,\eps,h) \ra \rho^c(p,\eps,h) = \rho(-p,\eps,-h) 
\ee
 
\subsection{Computation of a generic amplitude}

We have now all the elements necessary to compute 
a generic amplitude of the form ${\cal M} = tr[{\cal O}\Sigma]$. 
We therefore wish to proceed in a similar way as we did in the 
simple example of section~\ref{example}. Thus we can write 
the equivalence:

\be
u\bar u' \equiv \frac{u \bar u k\sla u' \bar u'}
{\sqrt{tr[u \bar u k\sla u' \bar u' k\sla]}}
\ee

where the 4-vector $k$ serves as a reference for the phase of 
the amplitude and the equivalence means ``up to a phase''. It 
is important to note at this stage that we can be led into trouble 
if an amplitude contains several identical particles. In this 
case we should have the same reference for the phase of two  
diagrams where two identical fermions are permuted, because of the 
possible interferences. In this case, this method shouldn't work 
because we have arbitrarily changed the relative phase between the 
diagrams. In this case, a covariant expression can be obtained 
through the computation of the square of the amplitude, which 
often means huge analytical expressions. We shall not discuss 
further this problem. Coming back to our single fermionic current, 
 we can generically write the $\Sigma$ matrix in the form: 

\be
\Sigma = \frac{\rho_u k\sla \rho_{u'}}{\sqrt{tr[\rho_u k\sla \rho_{u'} 
k\sla]}} \label{generic}
\ee

where one can replace $\rho_u$ and $\rho_{u'}$ by their expression 
given in formula~\ref{rho} for massive fermions or eq.~\ref{massless1} 
for massless fermions. Then, using eq~\ref{generic}, the amplitude 
can be written: 

\be
{\cal M} = \frac{tr[{\cal O}\rho_u k\sla \rho_{u'}]}
{\sqrt{tr[\rho_u k\sla \rho_{u'} k\sla]}} \label{amp}
\ee

and one has to compute two simpler traces, instead of one possibly 
huge trace if one wants to compute the square of the 
amplitude using eq.~\ref{ampl-sq}. As for the choice of $k^\mu$, we 
have seen in the simple example presented in section~\ref{example} 
that we may have some problems if this vector is not suitably 
chosen, especially in the massless case, because of a 
kinematic singularity appearing in the plane generated by the 
two momenta. And if one intends to implement this formula inside 
a Monte-Carlo program, where the events are generated randomly, 
it is better to avoid this kind of singularity to be sure that 
none of the momentum configurations will be close to the singularity. 
Therefore in the massless case, one may choose $k^\mu$ 
such that $k^2 = 1$, and in the massive case, the explicit 
computation of the normalization factor in eq.~\ref{amp} shows 
that one may preferably choose $k^\mu$ on the light cone. Since 
the explicit calculation of these traces can be done easily 
by some symbolic calculation programs, we shall let the 
reader do them if he is interested in such calculations, 
because it is often important to take advantage of the 
particular situations to choose $k^\mu$, or the spin axes properly.

 We may end this section by mentioning that we can also 
 choose a simpler operator than $k\sla$ to be inserted 
into the normalization factor when at least one of the fermions 
is massive. We can simply use the identity matrix instead 
of $k\sla$ and we get:

\be
{\cal M} = \frac{tr[{\cal O}\rho_u \rho_{u'}]}
{\sqrt{tr[\rho_u \rho_{u'}]}} 
\ee

 Which is simpler to compute, and shall not lead to kinematical 
singularities in most cases. Since the denominator can be 
of order $mm'$, it will be numerically better to use it for 
heavy fermions. 

\section{Conclusion}

In this paper, we have shown a quite general method for 
the calculation of Feynman amplitudes or its square with external 
fermions. The mass of the fermions can be of any value, even 
in the same fermionic line. The formulas given are also independent 
of the representation of the Dirac matrices, explicitly covariant 
and have a meaningful massless limit. The formalism 
is not exactly continuous in $m=0$ in the strict 
sense since there are some degrees of freedom that appear in the 
case $m=0$ that compensate the transverse degrees that are 
``frozen'' into the two helicity modes when $m\ra 0$. However,  
these transverse degrees of freedom are not observed for neutrinos 
and they are therefore not taken under consideration for these 
particles. For the study of some New Physics, 
it can be interesting to keep them. The other trouble with massless 
fermions comes when one computes a scattering amplitude 
with a current composed of two of these massless fermions. There 
can be some singularity in phase space for the normalization 
condition if one takes a fixed momentum for the reference phase, 
which can lay in the plane of the two external momenta 
for some kinematic configurations. This singularity does 
not exist for the computation of the square of the amplitude, 
which fortunately in the massless case may lead to 
expressions of reasonable size, contrary to the massive case. 

To obtain the general expression of an amplitude, we started 
from well known formulas giving the density matrices 
that we have redemonstrated in the most general case. 
Using this kind of demonstration, the possibility of a transverse 
polarization for massless fermions naturally emerges, and we 
have also characterized within this formalism the notion of charge 
conjugation and Majorana spinors. We expect that these results 
can be useful to get simpler analytic expressions for 
short Feynman amplitudes, or not-so-short amplitudes computed 
using symbolic calculation programs. We also expect a 
simpler implementation of fermions in Monte-Carlo programs, thanks 
to the fact that the formalism presented in this paper requires 
very few conventions, which was also one of the goals of this work.

\subsubsection*{Acknowledgments} 

I must thank Pr.~W.~Pezzaglia and Pr.~I.~Benn 
for their interesting answers to my questions, and P.~Overmann 
for his advice in the practical use of his symbolic 
calculation program ``M''.
 

\section*{Appendix A: some properties of the Dirac matrices}

 Suppose that we have one set of $\gamma^\mu$ matrices that obey 
the anticommutation rules. From this set of $4$ matrices, one can 
explicit a basis of ${\cal M}_4(C)$\footnote{${\cal M}_4(C)$ denotes 
the set of $4\times 4$ complex matrices throughout the paper.}, 
$\gamma^r$ ($r~\in~[\![1..16]\!]$) in this way: 

\be 
\gamma_r = {\gamma_0}^{\alpha_0(r)}{\gamma_1}^{\alpha_1(r)}
{\gamma_2}^{\alpha_2(r)}{\gamma_3}^{\alpha_3(r)}
\ee 

with $\alpha_\mu(r) = 0,1$. We may also sometimes denote 
$n_\gamma(r) = \alpha_0(r)+\alpha_1(r) +\alpha_2(r) + \alpha_3(r)$. 
When $n_\gamma(r)$ is even, the matrix is said to be in the 
even subalgebra of the Clifford algebra.  
This basis is chosen essentially for the proof of the theorem 
relating two different sets of Dirac matrices (see beyond), 
but most of the time one uses the more convenient basis 
($I=\gamma_{(r=1)}$, $\gamma^5 = i\gamma^0\gamma^1\gamma^2\gamma^3= 
i\gamma_{(r=16)}$, $\gamma^\mu$, $\gamma^\mu\gamma^5$, 
$\sigma^{\mu\nu} = \frac{i}{2}[\gamma^\mu,
\gamma^\nu]$)\footnote{Sometimes we will have to 
consider $\gamma_{(r=1)}$ which is just the identity, different from 
the usual $\gamma_{(\mu=1)}$. We will therefore write the ``$r=...$'' 
when necessary, in order to avoid some confusions about the 
meaning of the indices.}. 

With these definitions one can prove that $tr[\gamma_r] = 0$ 
if $\gamma_r$ is not the identity matrix ($r\ne 1$) and 
that $\gamma_p \gamma_q = \eta_{p,q} 
\gamma_{I(p,q)}$, where $\eta_{p,q} =\pm 1$, and more 
importantly, for a fixed $q$, $p\mapsto I(p,q)$ is a permutation 
of $[\![1..16]\!]$ (and similarly for a fixed $p$,  
$q\mapsto I(p,q)$ is also a permutation). Also $\eta_{p,q}$ depends 
only on the anticommutation relations and the metric convention, 
and not on the specific set (representation) of Dirac matrices 
chosen. Since we have also shown in the beginning of the paper 
that the only relevant signature is $(1,3)$, we also assume 
this metric to be chosen. We have also 
$tr[\gamma_i\gamma_j] = C_{ij}$, where the coefficients $C_{i,j}$ 
 vanish if and only if $i\ne j$.

\subsubsection*{Demonstration of the relation between different 
representations}

Suppose now that we have two sets of Dirac matrices $\gamma^\mu$ 
and ${\gamma'}^\mu$ that obey the anticommutation rules. As before, 
we can define a basis for the $4\times 4$ complex matrices 
${\gamma}_r$, and another similar basis ${\gamma'}_r$ defined in 
the same way but using  ${\gamma'}^\mu$ instead of $\gamma^\mu$. 
We shall also need another $4\times4$ complex matrix, $F$ upon 
which we do not impose any constraint at this time. Then we 
define the matrix $S$ as:

\be
S = \sum_{r=1..16} {\gamma'}_r F {\gamma_r}^{(-1)}
\ee 

We will show that $S$ must be $0$ or invertible, 
and then that in this latter case one has 
${\gamma'}^r = S\gamma^r S^{-1}$ (it is sufficient to check 
this property for $r\equiv \mu~\in~[\![0..3]\!]$). 

 The proof comes from the following relation: 
\bea
 S\gamma_t &=& \sum_{r=1..16} {\gamma'}_r F {\gamma_r}^{(-1)}\gamma_t 
\nonumber \\
&=& \sum_{s=1..16}{\gamma'}_{I(t,s)} F {\gamma_{I(t,s)}}^{(-1)}
\gamma_t~~~~~~~~~~~~(\Leftarrow~~~ r \ra I(t,s) ) \nonumber \\
&=& \sum_{s=1..16}{\gamma'}_{I(t,s)} F {\gamma_s}^{(-1)} 
{\gamma_t}^{(-1)}{\gamma_t} \eta_{t,s}~~~~~~~~~~~~~(\Leftarrow~~ 
def.~~of~\gamma_{I(t,s)}) \nonumber \\
&=& \sum_{s=1..16} \eta_{t,s}{\gamma'}_{I(t,s)} F {\gamma_s}^{(-1)} 
\nonumber \\
&=& \sum_{s=1..16} {\gamma'}_t {\gamma'}_s F {\gamma_s}^{(-1)}  
~~~~~~~~~~~~~(\Leftarrow~~ def.~~of~\gamma_{I(t,s)})\nonumber \\
&=& {\gamma'}_t S
\label{fondam-th}
\ena

This is the very fundamental property we need to prove our latest 
assertion. Now if $x$ is a four vector such that $Sx=0$, then 
every $\gamma_r x$ is in the kernel of $S$. It means that 
$Ker\, S$ is stable through the action of the whole algebra, that 
is to say every $4\times 4$ complex matrix (${\cal M}_4(C)$). If 
we consider an irreducible representation of the Dirac matrices, 
it implies that $Ker\, S$ must be an improper subspace, i.e. $0$ or 
the whole spinor space $C^4$. Therefore, $S$ is invertible or 
$0$. 
 Now, why is the four dimensional representation irreducible? 
Suppose that $K = Ker\, S$ is not $0$, and let $L$ be a subspace 
of $C^4$ such that $C^4 = K\oplus L$. Let ${\cal M}$ be an
endomorphism of $C^4$ such that the image of one (non-zero) vector 
in $K$ lies in $L$ (it only needs $L$ to be different from $0$). 
Now, the key point is that ${\cal M}$ can be decomposed on 
the $\gamma_r$ basis. Since $K$ is stable by all the $\gamma_r$, 
we get a contradiction and $L$ must be $0$, 
$\Rightarrow K =C^4 \Rightarrow S=0.~\Box$

  We have now demonstrated that two sets of Dirac matrices are 
related by an inner automorphism if we can find $S \ne 0$. And it is 
possible to find $F$ such that $S$ is not $0$. Otherwise, for  
all $x,y\,\in\,C^4$, posing $F=yx^\dagger$, one has 
$\sum_{r} (y^\dagger {\gamma'}_r y) x^\dagger {\gamma_r}^{(-1)} =0$, 
thus $\sum_{r} (y^\dagger {\gamma'}_r y) {\gamma_r}^{(-1)} =0$ and 
taking the trace, one obtains $y^\dagger{\gamma'}_{(r=1)} y = 
4y^\dagger y$ which cannot be $0$ if $F\ne 0$. So one can 
find $F$ such that $S$ is non-zero, then $S$ is invertible 
and the two representations are related through an inner 
automorphism.~$\Box$ 

\subsubsection*{Unitary representations}

 In this paragraph, we will show that one can construct a unitary 
representation of the Dirac matrices from any irreducible 
representation. 

 So, if we denote $H = \sum_{r} \gamma_r^\dagger\gamma_{r}$, 
one can use some arguments similar to the ones used 
in eq.~\ref{fondam-th} 
(with also the fact that $\eta_{p,q}^2=1$) to show that 
$\gamma_\mu^\dagger H\gamma_\mu =  H$. Now, it is clear that 
$H$ is hermitian definite and positive, which allows us to 
write it as a ``square'', i.e. $H=h^\dagger h$ and $h$ is 
invertible\footnote{$h$ is also unique if it is taken hermitian 
positive definite.}. Thus we have 
$\gamma_\mu^\dagger h^\dagger h\gamma_\mu =  h^\dagger h$ and therefore 
the representation given by $h\gamma_\mu h^{-1}$ is unitary.~$\Box$

\section*{Appendix B: mathematical proof of some elementary theorems}

In this appendix we will demonstrate two interesting properties. 
The second one is used in the body of this paper. The first one 
is rather trivial if one uses a specific representation 
for the Dirac matrices. The aim here is just to illustrate that most 
of the important properties within the Dirac formalism can be shown 
without the need of any specific representation for the $\gamma$ 
matrices. 

\noindent{\bf Theorem 1:} {\it Let $p$ be a non vanishing four 
momenta. 
We denote $m=\sqrt{p^2}$ if $p^2\geq 0$ and $m=i\sqrt{-p^2}$ 
if $p^2<0$. Let $c$ be a constant which can be set 
to $\pm 1$ (if the convention for the metric is ($+$\,$-$\,$-$\,$-$)), 
or $\pm i$ (if the convention for the metric is ($-$\,$+$\,$+$\,$+$), 
case which can be avoided if we replace $p\sla$ by $\tilde p$, 
as shown in the beginning of the paper). 
Then $Ker(p\sla-cmI) = Im(p\sla+cmI)$.}

{\bf Proof:} 
Let $M_4$ be the four dimensional Minkowski space, and 
$C^4$ a four dimensional complex vector space, in which 
we define spinors. \\
Since $p\sla^2 = c^2m^2I$, it is clear that 
$Im(p\sla+cmI) \subset Ker(p\sla-cmI)$.
In the case $m \neq 0$, if $u \in Ker(p\sla-cmI)$, then 
$u = (p\sla+cmI)(u/(2cm))$, and the theorem is proved. In fact, 
the eigenvalues of $p\sla$ are of order two, because $p\sla$ 
has at most 2 non-vanishing eigenvalues, and its trace must be 
zero. Thus $p\sla-cmI$ are two rank two matrices, 
and we recover the elementary result that the space of states for 
a spin~1/2 particle is of dimension 2. This is also true when 
$m=0$ as we shall see.\\

Now suppose that $m=0$. We know that $dim~Ker(p\sla) + 
dim~Im(p\sla) =4$, and we have seen that 
$Im(p\sla) \subset Ker(p\sla)$, and since 
$p^\mu$ is a non vanishing vector, $p\sla \neq 0$ and one has: 
$1 \leq dim~Im(p\sla) \leq dim~Ker(p\sla)$. Thus we have two 
possibilities: 
\begin{itemize}
\item{$dim~Im(p\sla) = 2 = dim~Ker(p\sla)$. Then the theorem is 
proved.}
\item{$dim~Im(p\sla) = 1$ and $dim~Ker(p\sla)=3$.}
\end{itemize}
 
We will therefore show that the second case is not possible. 
Let $u_i$, $i \in \{0,1,2,3\}$ be a basis of $C_4$, such that 
$\{u_1,u_2,u_3\}$ is a basis of $Ker(p\sla)$ and $u_1 = p\sla u_0$. 
The vector $p^\mu$ is of the form ${\bf p}(1,\vec n)$ 
(with ${\vec n}^2=1$). Let $\vec n'$ and 
$\vec n''$ two vectors such that $\vec n'$, $\vec n''$ and $\vec n$ 
form an orthonormal basis of $R^3$. 
Then if $q' = (0,\vec n')$ and $q'' = (0,\vec n'')$, 
one has $(q'\cdot p) = (q''\cdot p) = (q'\cdot q'')=0$, 
and ${q\sla'}^2 = 
{q'}^2 = \pm 1 = {q\sla''}^2$. Therefore, the eigenvalues of $q\sla'$ 
and $q\sla''$ are in the set $\{i,-i\}$ or $\{1,-1\}$ depending on the 
sign convention for the metric. \\

One has $u_1 = p\sla(u_0)$, then $q\sla' u_1 = p\sla (-q\sla' u_0)$, 
but if $dim~Im(p\sla) = 1$, we must have $q\sla' u_1 = \lambda' u_1$ 
with $\lambda' = \pm i$ (or $\pm 1$), and similarly for $q\sla'' u_1$. 
Now $q\sla' (q\sla'' u_1) = \lambda'\lambda'' u_1 = 
-{q\sla'' ({q\sla'} u_1)} = -\lambda'\lambda'' u_1$. This would 
imply $\lambda'\lambda'' u_1=0$,  which is impossible. 
Thus the only possibility is $dim~Im(p\sla) = 2 = dim~Ker(p\sla)$, 
which implies  $Im(p\sla) = Ker(p\sla)$, and the theorem is proved 
also for $m=0$.~$\Box$ \\

\noindent{\bf Theorem 2:} {\it Let $M$ be a $n\times n$ complex matrix. 
then one has:}
\be
rank(M)=1 \Leftrightarrow M\neq 0~and~\forall Q\in {\cal M}_n(C)~~
MQM = tr(MQ)M
\ee

{\bf Lemma:} {\it Let $M$ be a $n\times n$ complex matrix, such 
that $tr(M)\neq 0$. Then :}
\be
rank(M)=1 \Leftrightarrow M^2 = tr(M)M
\ee
\indent{\bf Proof:}  
  
If $M$ is a rank~1 matrix, then one can write $M=xy^\dagger$ where 
$x$ and $y$ are two complex vectors. Since we have $tr(M) = 
y^\dagger\cdot x$, one has $M^2=tr(M)M$.\\

Conversely, since $tr(M)\neq 0$ then $X\wedge (X-tr(M)) =1$, and 
the condition $M(M-tr(M)) =0$ implies that 
$C^n = Ker(M)\oplus Ker(M-tr(M)I)$ ($n=4$), thanks to the 
kernel decomposition theorem. Thus, using a basis compatible 
with this decomposition, $M$ can be written in the form:

\be M=P^{-1}\left( \begin{array}{cccc} 0&&0& \\
 &tr(M)&& \\ 0&&\ddots& \\ &&&tr(M) \end{array} \right)P
\ee

Where $P$ is an invertible matrix. It leads 
to $tr(M) = dim~Ker(M-tr(M)I)\times tr(M)$, thus 
$dim~Ker(M-tr(M)I) =1\Rightarrow rank(M)=1$.~$\Box$ \\

\indent{\bf Proof of the theorem:}

If $rank(M)=1$, then $\forall Q,~rank(MQ)=0~or~1$, and 
applying the same reasoning as in the lemma, we easily conclude that 
$MQM = tr(MQ)M$.\\

Conversely, if $\forall Q,~tr(MQ)=0$ then $M=0$. Therefore, in 
our case, we can find $Q$ such that $tr(MQ)\neq 0$. And 
since $Q\mapsto tr(MQ)$ is a continuous function, and  
the group of invertible matrices is dense in ${\cal M}_n(C)$, 
we can choose $Q$ invertible. Then, multiplying on the right side 
by the matrix $Q$ we get $(MQ)^2=tr(MQ)MQ$ and the lemma tells us 
then that $MQ$ is a rank~1 matrix, and thus $M=(MQ)Q^{-1}$ is also 
a rank~1 matrix.~$\Box$

\noindent{\bf Corollary:} {\it Let $M$ be a $n\times n$ complex 
matrix ($M\ne 0$). We have the following equivalence:}
\be
rank(M)=1 \Leftrightarrow \forall (Q,Q')~~ tr(MQMQ')=tr(MQ)tr(MQ')
\ee
\indent{\bf Proof:} 

If one knows that $M$ is a rank 1 matrix, the conclusion is a direct 
consequence of theorem~2. Conversely, if we know that 
$\forall (Q,Q')~~tr(MQMQ')=tr(MQ)tr(MQ')$, then the linear form 
$Q'\mapsto tr(MQMQ')-tr(MQ)tr(MQ')$ vanishes, which implies 
that $MQM =tr(MQ)M$ for every matrix $Q$. Then, the theorem 2 
tells us that $M$ is a rank~1 matrix.~$\Box$


\end{document}